\documentclass[%
 aip,
 amsmath,amssymb,
 reprint,%
]{revtex4-1}

\usepackage{graphicx}
\usepackage{dcolumn}
\usepackage{bm}

\usepackage[utf8]{inputenc}
\usepackage[T1]{fontenc}
\usepackage{mathptmx}

\begin{document}

\preprint{AIP/123-QED}

\title[On the illposedness and stability of the relativistic heat equation]{On the illposedness and stability of the relativistic heat equation}

\author{A. L. García-Perciante}
 \email{algarcia@correo.cua.uam.mx}
 \affiliation{Depto. de Matemáticas Aplicadas y Sistemas,\\
Universidad Autónoma Metropolitana-Cuajimalpa.\\
Cuajimalpa de Morelos (05348), Ciudad de México.}
\author{O. Reula}
 \email{reula@famaf.unc.edu.ar}
\affiliation{Facultad de Matemática, Astronomía, Física y Computación,\\ Universidad Nacional de Córdoba. Instituto de Física Enrique Gaviola, CONICET.\\ Ciudad Universitaria (5000), Córdoba, Argentina.}

\date{\today}

\begin{abstract}
In this note we analyze, in terms of a simple example, the incompatibility of parabolic evolution and general covariance. For this we introduce a unit time-like four-vector and study the simplest heat flux equation with respect to it. In cases where this vector field is surface forming then the local high wave number limit shows well posedness, but as soon as that property is lost the Cauchy problem becomes ill-posed. We also discuss how the Maxwell-Cattaneo type modification of the system renders it well posed and link the amplitude of the modification, which is related to the so-called second wave speed of the system, to the size of the failure of surface orthogonality.
\end{abstract}

\maketitle

\section{\label{sec:level1}Introduction }

In Galilean space-times we have systems of equations describing dissipative
fluids which lead to a mixture of wave propagation and dissipative damping.
They constitute hyperbolic-parabolic systems and are built in general
following the basic example of the Navier-Stokes system. In these
theories there is a field representing the velocity of the fluid plus
some other quantities representing densities, of energy and particle number
and then constitutive relations linking several dissipative fluxes
to derivatives of some of the above fields. Together with energy,
momentum, and particle number conservation they constitute a closed
evolution system.

It is well known that the heat, or diffusion, equation which is of
parabolic nature is not adequate in general since it allows for arbitrarily
large propagation speeds. In a relativistic scenario it is usually
replaced by a Cattaneo \cite{Cattaneo} type equation, a procedure which renders the
system hyperbolic by introducing, without a formal derivation, a time
derivative term in Fourier's law. This promotes the heat flux to a
dynamical state variable and the corresponding constitutive equation
to a hydrodynamic equation instead of the constraint relation status
that it possesses in the conventional theory. 
Moreover, the Cattaneo equation can be thought of as the precursor of the widely used extended thermodynamics \cite{jou} and second order theories \cite{Israel} which have been thoroughly analyzed in a formal mathematical fashion \cite{geroch,GEROCHann}.
This topic has been widely addressed theoretically,
experimentally and numerically. However, the illposedness and stability
of both parabolic and hyperbolic versions of the heat equation is
still obscure. 
Moreover, the relevant role of the space-time representation
in the nature of the system has been practically overlooked.

Other models have been proposed in order to attain the desired consistency between special relativity and non-equilibrium thermodynamics (see for example \cite{Li} and references cited therein). However, such works focus primarily in the causal nature of the heat conduction law and do not analyze the system in the more complete, covariant nature.

In this short note, we analyze the dispersion relation that one is lead to in the case of a Fourier heat conduction law, as well as including a heat relaxation term following Cattaneo's idea, in a covariant fashion by considering two cases. In the case the assumed four-velocity vector is surface forming one can establish a 3+1 decomposition where the normal to the surfaces is precisely that four-velocity for all members of the foliation. In this case the corresponding Cauchy problem is well posed, that is solutions to the future of a given surface depend continuously on the initial data. In the generic case, where the four-velocity is not surface forming, no such foliation exists. Thus a Cauchy problem has to be defined using an arbitrary foliation, starting from and arbitrary initial surface. The parabolic nature of the equations imply the existence of arbitrarily large propagation speeds as the wave number goes to infinity, but the surfaces under consideration can not be aligned to these surfaces and so perturbations influence the past of near-by points. This inconsistency is reflected in arbitrary blows ups of perturbations as wave numbers increase. Thus, apart from very particular, highly symmetric, irrotational situations the equations are ill-posed.

In order to address the point described above in a clear and brief
fashion, we have organized the rest of this note as follows. In Section
II we state the problem by writing the heat equation as a system of
partial differential equations both in a Galilean and covariant fashion
and study the corresponding dispersion relations for plane waves solutions
in a linear approximation firstly in a comoving frame, assuming one
can define such frame (i. e. the hydrodynamic velocity forms surfaces),
and secondly following Hiscock et. al. \cite{HL}, considering a boost to an arbitrary frame. The procedure
is repeated in Sect. III but for a system where the Fourier law is replaced
by a Cattaneo-type dynamic equation for the heat flux. The conclusions
of the analysis and final remarks are included in Sect. IV.
\section{\label{sec:level}The parabolic heat equation}
To illustrate the problem at hand it suffices to consider the following simple Galilean system (for a discussion on Galilean space-times see for example \cite{weyl}):
\begin{equation}
\frac{DT}{Dt}=-\frac{1}{nc_{n}}h^{ab}\nabla_{a}q_{b},\label{1}
\end{equation}
\begin{equation}
q_{a}=-\kappa\tilde{\eta}_{a}{}^{b}\nabla_{b}T,\label{1.1}
\end{equation}
where $\frac{D}{Dt}=\left(\frac{\partial}{\partial t}+u^{a}\frac{\partial}{\partial x^{a}}\right)$
is the total (material) derivative, $c_{n}=\left(\frac{\partial e}{\partial T}\right)_{n}$
is the specific heat for constant particle number, $\kappa$ is the thermal conductivity coefficient, and $\tilde{\eta}^{ab}$ the euclidean flat three-metric.
The covariant, relativistic, version of the system in Eq. (\ref{1})
is given by:
\begin{equation}
    u^{a}\nabla_{a}T= -\frac{1}{nc_{n}}h^{ab}\nabla_{a}q_{b},\label{2}
\end{equation}
\begin{equation}
q_{a}=-\kappa h_{a}^{b}\nabla_{b}T.
\end{equation}
Where now $u^{a}$ denotes the fluid's four velocity, and $h^{ab}=\eta^{ab}+u^{a}u^{b}$,
where $\eta^{ab}$ is the flat four-metric with a $(-1,1,1,1)$ signature and a normalization, $u^{a}u_{a}=-1$. 
\subsection{\label{sec:level21}Stability in the surface forming case.}
If the four-vector $u^{a}$ is surface forming, that is, we can choose global surfaces whose tangent space are everywhere normal to $u^{a}$, then we can use one of them to pose a Cauchy problem, namely to give the value of $T$ at it and obtain a unique solution to the future of such surface. If we were in such a case we would see that perturbations propagate with arbitrary high speeds and so given any perturbation initially of compact support its solution would spread to all space-time for any arbitrarily short time to the future of our initial slice. This is in contradiction with the axiom of maximal propagation speed (the speed of light) assumed in general relativity but actually is mathematically consistent and results in a causal propagation where causal now is in the pre general-relativity sense. Namely instead of maximal propagation cones we have maximal propagation hypersurfaces. In the limit of infinite frequency perturbations propagate along them.

Indeed, to see this look at the principal symbol of the system and
analyze high frequency solutions. This suffices to assert whether
the system is well posed or not. Substituting time derivatives ($u^{a}\nabla_{a}$)
by $s$ and space derivatives by $ik^{a}$, with $u^{a}k_{a}=0$ we
get,
\begin{equation}
sT=-\frac{i}{nc_{n}}k^{b}q_{b},
\end{equation}
\begin{equation}
q_{a}=-i\kappa k_{a}T,\qquad\kappa>0,
\end{equation}
out of which we get the associated matrix,
\begin{equation}
M=\left(\begin{array}{cc}
s & \frac{i}{nc_{n}}\\
\kappa ik^{2} & 1
\end{array}\right),\label{8}
\end{equation}
and from its determinant the dispersion relation (where we have defined
$\alpha=\frac{\kappa}{nc_{n}}$) 
\begin{equation}
s+\alpha k^{2}=0.\label{9}
\end{equation}
Thus, the system is stable and fluctuations decay to the future with a characteristic time $\left(\alpha k^{2}\right)^{-1}$. 
\subsection{\label{sec:level22}The general, not-surface forming case.}

But what happens when the four-velocity $u^{a}$ is not surface forming?
To see this imagine in the previous case we take any space-like surface
as Cauchy surface to start our evolution. We are in trouble: the perturbations
are still traveling with arbitrarily large speeds but since our initial
surface is different that the maximal propagation surface, signals
would propagate to the future or the past of it, thus spoiling the
unique solution we are seeking to construct.

To see this in formulae consider now the construction with a different
vector, that is, we choose a four-vector $n^{a}$ as being surface
forming and seek to pose the Cauchy problem along these hypersurfaces:
Writing, 
\begin{equation}
g^{ab}=e^{ab}-n^{a}n^{b},\quad n^{a}n_{a}=-1,\quad e^{ab}n_{a}=0,
\end{equation}
and,
\begin{equation}
u^{a}=\gamma\left(n^{a}+\beta^{a}\right),\quad\beta^{a}n_{a}=0,
\end{equation}
where $\gamma$ is the corresponding Lorentz factor for a boost with
relative velocity $\beta^{a}$. In such frame, the differential operator
for plane-wave solutions is given by: 
\begin{equation}
\tilde{\nabla}_{a}=-sn_{a}+ik_{a},\,\text{with}\,k^{a}n_{a}=0,\label{b2}
\end{equation}
with the following space, time and hydrodynamic velocity projections,
\begin{equation}
n^{a}\tilde{\nabla}_{a}=s,\,\,k^{a}\tilde{\nabla}_{a}=ik^{2},\,\,u^{a}\tilde{\nabla}_{a}=\gamma\left(s+ik_{\beta}\right),
\end{equation}
where we have defined $k_{\beta}=k_{a}\beta^{a}$. In order to obtain
$h^{ab}\tilde{\nabla}_{b}$ we write 
\begin{equation}
h^{ab}=e^{ab}-n^{a}n^{b}+u^{a}u^{b},
\end{equation}
and thus 
\begin{equation}
h^{ab}\tilde{\nabla}_{a}=ik^{b}-sn^{b}+\gamma\left(s+ik_{\beta}\right)u^{b}.\label{haboost}
\end{equation}
\begin{equation}
\epsilon\gamma\left(s+ik_{\beta}\right)q_{a}+q_{a}=-\kappa h_{a}^{b}\nabla_{b}T.
\end{equation}
Finally, the system in Eq. (\ref{1}) within this representation ($\nabla\rightarrow\tilde{\nabla}$
) is given by 
\begin{equation}
\gamma\left(s+ik_{\beta}\right)T= -\frac{1}{nc_{n}}\left(ik^{b}-sn^{b}+\gamma\left(s+ik_{\beta}\right)u^{b}\right)q_{b},\label{2-2}
\end{equation}
\begin{equation}
q_{a}=-\kappa\left(ik_{a}-sn_{a}+\gamma\left(s+ik_{\beta}\right)u_{a}\right)T,
\end{equation}
which, considering $T$ and the scalar $\left(ik^{a}-sn^{a}\right)q_{a}$
as independent variables, results in the following associated matrix
\begin{equation}
M=\left(\begin{array}{cc}
\gamma\left(s+ik_{\beta}\right) & \frac{1}{nc_{n}}\\
-\kappa\left(s^{2}+k^{2}-\gamma^{2}\left(s+ik_{\beta}\right)^{2}\right) & 1
\end{array}\right),
\end{equation}
and dispersion relation
\begin{equation}
\left(\alpha\beta^{2}\gamma^{2}\right)s^{2}+\left(2ik_{\beta}\gamma\alpha-1\right)\gamma s-\left(\gamma^{2}\alpha k_{\beta}^{2}+\gamma ik_{\beta}+\alpha k^{2}\right)=0.\label{disp}
\end{equation}
Taking $\beta\rightarrow0$ one gets $\gamma s+\alpha k^{2}=0$, which
leads to the decay obtained in the previous section. But for any non-zero
boost, Eq. (\ref{disp}) has at least one root in the right half of
the complex plane (using a generalization of RH criterion). In order
to see this one can normalize the equation as, 
\begin{equation}
s^{2}+\left(p_{1}+iq_{1}\right)s+\left(p_{2}+iq_{2}\right)=0,\label{pes}
\end{equation}
with 
\begin{equation}
p_{1}=-\left(\alpha\beta^{2}\gamma\right)^{-1},\quad q_{1}=\frac{2k_{\beta}}{\beta^{2}},\nonumber 
\end{equation}
\begin{equation}
p_{2}=-\frac{1}{\beta^{2}}\left(k_{\beta}^{2}+\frac{k^{2}}{\gamma^{2}}\right),\quad q_{2}=-\frac{k_{\beta}}{\alpha\beta^{2}\gamma}.\nonumber 
\end{equation}
The criterium for all roots to have negative real parts is given by
$p_{1}>0$ and $q_{2}^{2}<p_{1}\left(p_{1}p_{2}+q_{1}q_{2}\right)$,
which is clearly not satisfied. Thus, there is at least one root with
positive real part. Indeed, the real part of the roots is given by
\begin{equation}
\mathrm{Re}\left(s_{1,2}\right)=\frac{1}{2\beta^{2}\gamma\alpha}\left\{ 1\pm\mathrm{Re}\left(\sqrt{\zeta}\right)\right\}, 
\end{equation}
where 
\begin{equation}
\zeta=1+4\alpha^{2}\left(k^{2}\beta^{2}-k_{\beta}^{2}\right)-4\alpha i\left(\frac{k_{\beta}}{\gamma}\right).
\end{equation}
Defining, $k_{\beta}=k\beta\chi$ ($-1\leq\chi\leq1$), one can write
\begin{equation}
\begin{split}
\mathrm{Re}\left(\sqrt{\zeta}\right) & =\frac{1}{\sqrt{2}}\left\{ 1+\left(2\alpha k\beta\right)^{2}\left(1-\chi^{2}\right)\right.\\
 & \left.+\sqrt{\left[1+\left(2\alpha k\beta\right)^{2}\left(1-\chi^{2}\right)\right]^{2}+\left(4\alpha k\beta\gamma\chi\right)^{2}}\right\} ^{1/2}
\end{split},
\end{equation}
which, for large values of $k$ and $\chi\neq0$ leads to
\begin{equation}
\mathrm{Re}\left(s_{1,2}\right)=\frac{1}{2\beta^{2}\gamma\alpha}\left\{ 1\pm2\alpha k\beta\sqrt{\left(1-\chi^{2}\right)}\right\}. \end{equation}
Notice that considering $k^{2}\beta^{2}=k_{\beta}^{2}$ (the boost is in the direction $k$) the expression simplifies to 
\begin{equation}
\mathrm{Re}\left(s_{1,2}\right)=\frac{1}{2\alpha\beta^{2}\gamma}\left\{ 1\pm\frac{1}{\sqrt{2}}\sqrt{1+\sqrt{1+\left(\frac{4\alpha k\beta}{\gamma}\right)^{2}}}\right\},
\end{equation}
which clearly shows that one mode decays and the other one grows for any value of $k$. In this case and for large values of $k$, the growth rate of the instability increases as $\sqrt{k}$: 
\begin{equation}
\mathrm{Re}\left(s_{1,2}\right)\sim\pm\sqrt{\frac{k}{2\alpha\gamma^{3}\beta^{3}}}.
\end{equation}
For general boosts, not necessarily in the direction of $\beta^{a}$,
the decay is even more prominent as it grows linearly with the wave frequency. 

Thus we conclude that this systems are not only unstable, but actually
ill-posed: by taking a sequence of initial data of bounded norm, but
of higher and higher frequency one can get solutions which grow arbitrarily
fast no matter how small the time interval is taken. Thus, the map from initial data to solutions is not continuous. Arbitrarily close to analytic solutions there are perfectly nice initial data points which diverge arbitrarily fast from the analytic one. Outside the very
particular cases of irrotational fluids these type of theories are
thus useless.

\section{Hyperbolizing fluids}
A simple way of rendering the above system well posed is to introduce
a flow derivative in the Fourier law. In its simplest version this
is the Maxwell-Cattaneo \cite{Cattaneo} equation, 
\begin{equation}
\epsilon u^{b}\nabla_{b}q_{a}+q_{a}=-\kappa h_{a}^{b}\nabla_{b}T\qquad\kappa>0\label{10}
\end{equation}
where $\epsilon$ is supposedly a small parameter.

As mentioned above, these ideas have been generalized and improved in order to construct causal theories for relativistic gases \cite{jou,geroch,GEROCHann,GL}. Equation (\ref{10}) is here considered only as a simple example in order to explore the mechanism through which the inclusion of the additional relaxation term modifies the stability and ill-posedness of the system. In the following
subsections, a procedure analogous to the one shown above is carried
out in order to analyze the system given by Eqs. (\ref{2}) and (\ref{10}).
\subsection{Stability in the surface forming case}
Equation Eq. (\ref{10}) in the fluid's frame can we written as
\begin{equation}
\left(s\epsilon+1\right)q_{a}+\kappa ik_{a}T=0\label{70}
\end{equation}
which, upon contraction with $k^{a}$ yields
\begin{equation}
\left(s\epsilon+1\right)k^{a}q_{a}+i\kappa k^{2}T=0
\end{equation}
In this case, the associated matrix for the longitudinal modes (the transverse components lead to trivial, stable, solutions) is given by 
\begin{equation}
M=\left(\begin{array}{cc}
s & \frac{i}{nc_{n}}\\
\kappa ik^{2} & 1+s\epsilon
\end{array}\right)\label{12-1}
\end{equation}
and the corresponding dispersion relation
\begin{equation}
\epsilon s^{2}+s+\alpha k^{2}=0\label{13-1}
\end{equation}
Since all coefficients in Eq. (\ref{13-1}) are positive, both roots
of the dispersion relation lay on the left half of the complex plane.
More precisely
\begin{equation}
s_{\pm}=\frac{1}{2\epsilon}\left(-1\pm\sqrt{1-2\alpha\epsilon k^{2}}\right)
\end{equation}
and fluctuations decay with a characteristic time $\left(2\epsilon\right)^{-1}$.
In the limit $\epsilon\ll1$, $s_{+}=-\frac{1}{\epsilon}$ and $s_{-}=-\alpha k^{2}$.
So we have two modes, the usual one and another which decays very
quickly. Indeed in the limit $\epsilon \to 0$ we loose a root, it goes to infinity. 
So for small $\epsilon$ it is arbitrarily big. The extra mode that this modification introduces is very short lived and does not affect substantially the evolution. Its initial imprint is quickly washed away and normal diffusion follows.
\subsection{The general, not-surface forming case}
For the non surface forming case we shall need a minimum modification of the equations,
this should be proportional to the magnitude of the failure of the velocity vector to be surface orthogonal.  
Defining $\lambda=\epsilon-\alpha \beta^2$, the dispersion relation in this case can be written as in Eq. (\ref{pes}),
where now
\begin{equation}
p_{1}=\left(\gamma\lambda\right)^{-1},\quad q_{1}=\frac{2k_{\beta}\left(\epsilon-\alpha\right)}{\lambda}\nonumber 
\end{equation}
\begin{equation}
p_{2}=\frac{\left(\gamma k_{\beta}^{2}\left(\alpha-\epsilon\right)+\alpha k^{2}\right)}{\gamma\lambda},\quad q_{2}=\frac{k_{\beta}}{\gamma\lambda}\nonumber 
\end{equation}
which reduces to Eq. (\ref{disp}) for $\epsilon=0$. The criteria
for stability, $p_{1}>0$ and $q_{2}^{2}<p_{1}\left(p_{1}p_{2}+q_{1}q_{2}\right)$,
is then
\[
\lambda>0,\quad\mathrm{and}\quad k_{\beta}^{2}<k^{2}.
\]
Notice that the second condition is satisfied as long as $k\neq0$
and $\chi\neq1$, and thus in such cases stability is guaranteed taking 
$\epsilon$ large enough such that $\epsilon>\alpha\beta^{2}$. 

The case of homogeneous perturbations ($k=0$) leads to one pure imaginary
root. The real part of the other (complex) root is $\left(-\gamma\lambda\right)^{-1}$
which coincides with the stability criteria stated above. On the other hand, for $k\neq0$, the real part of the roots can be written as
\begin{equation}
\text{Re}\left(s_{1,2}\right)=-\frac{1}{2\lambda\gamma}\left[1\pm\text{Re}\left(\sqrt{\eta+i\nu}\right)\right]\label{w}
\end{equation}
where, for $k_{\beta}=k\beta$, one has
\begin{equation}
\eta=1-4\frac{\alpha\epsilon}{\gamma^{2}}k^{2},\,\,\nu=-\frac{4\alpha\beta k}{\gamma}
\end{equation}
For large values of $k$, Eq. (\ref{w})
leads to
\begin{equation}
\text{Re}\left(s_{1,2}\right)\sim-\frac{1}{2\lambda\gamma}\left(1\pm\beta\sqrt{\frac{\alpha}{\epsilon}}\right)\label{eq:a}
\end{equation}
For the stable modes ($\lambda>0$) one obtains $\text{Re}\left(s_{1,2}\right)<0$, 
while in the unstable case one has ($\lambda<0$) 
\begin{equation}
\text{Re}\left(s_{1,2}\right)=-\frac{1}{2\lambda\gamma}\left(1\pm\beta\sqrt{\frac{\alpha}{\epsilon}}\right)\gtrless0\label{eq:a-1}
\end{equation}
Notice that the unstable mode is bounded and
approaches a constant value for large $k$. This implies that the
problem in this case is unstable but still well posed. In the general case, considering once again $k_{\beta}=k\beta\chi$, one has $\nu=-4\alpha k\beta\chi/\gamma$ and
\begin{equation}
\eta=1+4\alpha k^{2}\left[\alpha\beta^{2}\left(1-\chi^{2}\right)+\epsilon\left(\beta^{2}\chi^{2}-1\right)\right]\nonumber
\end{equation}
In this case, one obtains for large values of $k$:
\begin{equation}
\text{Re}\left(s_{1,2}\right)\sim-\frac{1}{2\lambda\gamma}\left\{ 1\pm\frac{\beta\chi}{\gamma}\sqrt{\frac{\alpha}{\lambda+\beta^{2}\chi^{2}\left(\frac{\alpha}{\gamma^2}-\lambda\right)}}\right\} <0\label{t}
\end{equation}
for $\lambda>0$, and for $\lambda<0$
\begin{equation}
\text{Re}\left(s_{1,2}\right)\sim-\frac{1}{2\lambda\gamma}\left\{1\pm k \sqrt{4\alpha\left[\beta^{2}\chi^{2}\left(\lambda-\frac{\alpha}{\gamma^2}\right)-\lambda\right]}\right\}\gtrless 0 
\end{equation}
Thus, the unstable mode grows linearly with $k$ for large
values of $k$, implying that the system is ill-posed.

Notice that in the stable case, one the modes features a very rapid decay. Indeed, for small values of $\lambda$, form Eq. (\ref{t}) one has
that $\text{Re}\left(s_{1}\right)$ is proportional to $\lambda^{-1}$, while $\text{Re}\left(s_{2}\right)$ is independent of $\epsilon$. Thus,  the solution decays to a sort of diffusion regime, but there isn't such regime, for things can not diffuse to arbitrary speeds. This has been studied, in some cases \cite{Nagy,Kreiss}

\subsection{How large does Catteno's term need to be?}

Cattaneo's term has to be larger than $\beta^2 \alpha$, so the question is, given some vector field $u_a$ in some space-time, can we find a space-like foliation such that the difference between $u_a$ and its normal $n_a$ is the smallest (presumably in pointwise norm) and how big is this difference? 

In principle, since $\beta <1$ we could take just this upper limit, which would give the speed of light as propagation speed of this new mode. But in many cases that might not be necessary and we could choose smaller speeds.

The measure of the local failure of surface forming for a given one-form field is given by the twist form:

\[
w_{abc} := u_{[a} \nabla_{b} u_{c]}.
\]
Notice that this is independent of any metric or torsion free connection. If $w_{abc}$ is different form zero it means that there is no pair of functions $(f, \tau)$ such that $f u_a = \nabla_a \tau$. 
So in principle one should be able to estimate the norm of $\beta_a = \frac{1}{\gamma}u_a - n_a$ in terms of some norm in $w_{abc}$. This is a very difficult task since the problem is of a global nature. Consider, for instance the case where we take $u_a = \gamma \left(1, 0, \beta_0,0\right) $ in Minkowski space-time in polar coordinates, $\left(t,r,\phi,z\right)$ with $\beta_0$ a constant smaller than unity. Here $\gamma= \left(1-\beta_0^2/r^2\right)^{-\frac{1}{2}}$.
This form is clearly non-surface-forming: if we follow the normal planes along $\phi$ we don't arrive at the same place after a turn around the $z$ axis. Nevertheless locally it is possible and the only place were $w_{abc}$ in non-vanishing is at the origin. So the bound on $\beta$ in terms of $w$ can not be local. If one relaxes $\beta_0$ to be a function of $r$, and writes for simplicity $u_a=\gamma\left(1,0,\beta_0(r)r,0\right)$ then one can have it to vanish near the origin (and so $w$). In this case the bound is given by, 
\[
|\beta_0| \leq \frac{1-e^{-r\max{|w|}}}{1+ e^{-r\max{|w|}}}.
\]

This follows from the fact that the only non-vanishing component of $w$ is $w_{tr\phi} = \gamma^2 \partial_r(\beta_0 r)$. 

As another illustrative example, one can allow a modulation on $\phi$ such that one can turn around the circle and correctly paste the planes. In this case the lack of surface forming comes from a lack of suitably in the $r$ component. Indeed, taking for simplicity the limit of small speeds, with $u_a = \left(1,0,\beta_0\left(r\right)r \cos\left(\phi\right),0\right)$, it can be seen that surfaces which glue well in the $\phi$ direction are possible, namely those given by $\tau = t + \beta_0\left(r\right) r \sin\left(\phi\right)$. However, they generate an $r$ component, $\partial_r \tau = \partial_r (\beta_0 r) \sin(\phi)$, which leads to the same bound as above (in the limit of small velocities). Thus one can see that is an intriguing problem where local PDE theory has to be analyzed in a global context.

\section{Final remarks}
In this short note, we have punctually addressed the question of whether a covariant heat equation, derived from either a Fourier or Cattaneo type relation with the help of an external four-vector field, leads to a physically sound theory. In particular, both schemes were thouroughly inspected for stability and well-posedness, using a method analogous to the one applied in Ref. \cite{HL}, in two relevant scenarios: when the fluid's velocity forms surfaces and when phenomena such as rotation does not allow for a particle frame to be utilized as a base for space-time 3+1 decomposition. The results are summarized in the following table.

\begin{table}[h!]
\begin{tabular}{|c|c|c|}
\hline
\multicolumn{1}{|l|}{} & Comoving frame                          & Boosted frame                                                                                                                                                                              \\ \hline
Fourier                & \begin{tabular}[c]{@{}c@{}}stable \\ well-posed \end{tabular}                      & unstable, ill-posed                                                                                                                                                                        \\ \hline
Cattaneo               & \begin{tabular}[c]{@{}c@{}}stable \\ well-posed \end{tabular} & \begin{tabular}[c]{@{}c@{}} stable for large $\epsilon$ \\ 
\begin{tabular}[c]{@{}c@{}}unstable \\ for small $\epsilon$ \end{tabular} 
$
\begin{cases}
\text{well-posed if }|\chi|=1   &\\ 
\text{ill-posed if }|\chi|\neq 1&
\end{cases}
$
\end{tabular} \\ \hline
\end{tabular} \label{table}
\end{table}

Notice that stability in the general case depends on the magnitude of the unspecified parameter $\epsilon$. Moreover, one could argue that under a close to equilibrium assumption, such parameter shall remain small when compared to the parameters of the system. However, the behavior of the perturbations depend strongly on the ratio $\frac{\epsilon}{\alpha \beta^2}$ and in particular are only well behaved when such quantity exceeds unity. In this sense it is worthwhile to notice that the kinetic theory of gases presents, at least within the standard Chapman-Enskog program, some inconsistencies in the derivation of such term \cite{JSP15} and thus, eventhough phenomenologically one can fix the value of $\epsilon$ arbitrarily, the soundness of such theory intuitively relies on considering only small values for it. Moreover, one can show that in Knudsen parameter expansion, the relaxation term included in the Cattaneo equation, is of higher order in such parameter and thus belongs in the Burnett regime equations. On the other hand, implementing a moment expansion one can justify  the structure of Eq. (\ref{10}) from microscopic grounds, however as is argued in Ref. \cite{doc}, the resulting transport equations feature terms of different order and require the introduction of an ordering scheme, in which the inconsistency exhibited in Ref. \cite{JSP15} will once again emerge.
\begin{acknowledgments}
O.R. acknowledges financial support from Conicet, SeCyT-UNC and MinCyT-Argentina.
\end{acknowledgments}
\nocite{*}
\bibliography{aipsamp}
\end{document}